\documentclass[]{spie} 

 \pdfoutput=1
\usepackage{amsmath,amsfonts,amssymb}
\usepackage{graphicx}
\usepackage[colorlinks=true, allcolors=blue]{hyperref}

\title{ALIOLI: Adaptive and Lucky Imaging Optics Lightweight Instrument}

\author[a,b]{Sergio Velasco}
\author[a,b]{Roberto L. L\'opez}
\author[a,b]{Alejandro Oscoz}
\author[a,b]{Carlos Colodro-Conde}

\affil[a]{Instituto de Astrof\'isica de Canarias, c/ V\'ia L\'actea s/n, La Laguna, Tenerife E-38205, Spain.}
\affil[b]{Departamento de Astrof\'isica, Universidad de La Laguna, La Laguna, Spain.}

\authorinfo{Further author information: (Send correspondence to S.V.)\\S.V.: E-mail: svelasco@iac.es}

\begin{document} 
\maketitle

\begin{abstract}
As a consequence of the evolution in the design and of the modularity of its components, AOLI for the William Herschel Telescope (WHT 4.2m) is much smaller and more efficient than its previous designs. This success has leaded us to plan to condense it even more to get a portable and easy to integrate system,  ALIOLI (Adaptive and Lucky Imaging Optics Lightweight Instrument). It consists of a DM+WFS module with a lucky imaging science camera attached.
ALIOLI is an AO instrument for the 1-2m class telescopes which will also be used as on-sky testbench for AO developments. Here we describe the setup to be installed at the 1.5m Telescopio Carlos Sánchez (TCS)  at the Spanish Observatorio del Teide (Tenerife, Canary Islands).
\end{abstract}


\section{The instrument}
\label{sec:intro} 

ALIOLI is the natural step to take after the successful AOLI\cite{Velasco2018a, VelascoSEA} instrument at WHT.

\begin{figure} [ht!]
\begin{center}
\includegraphics[height=9cm]{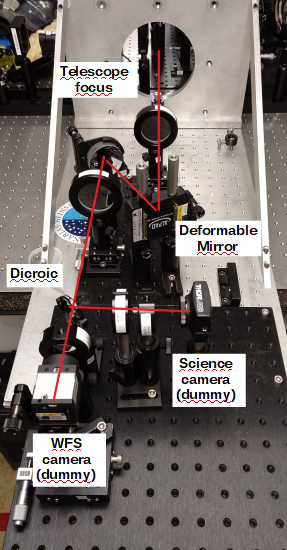}
\end{center}
\caption[labels] 
   { \label{fig:alioli_labels} 
View of the instrument from the back with the ray tracing superimposed. }
\end{figure} 

The development of AOLI during its AIV\cite{MartaSPIE, CraigSPIE} and commissioning phases\cite{Velasco2016} into a lighter and more versatile instrument \cite{RobertoSPIE}, gave up ALIOLI as the best solution to perform Adaptive Optics at medium-size telescopes.

\begin{figure} [ht!]
\begin{center}
\includegraphics[height=5cm]{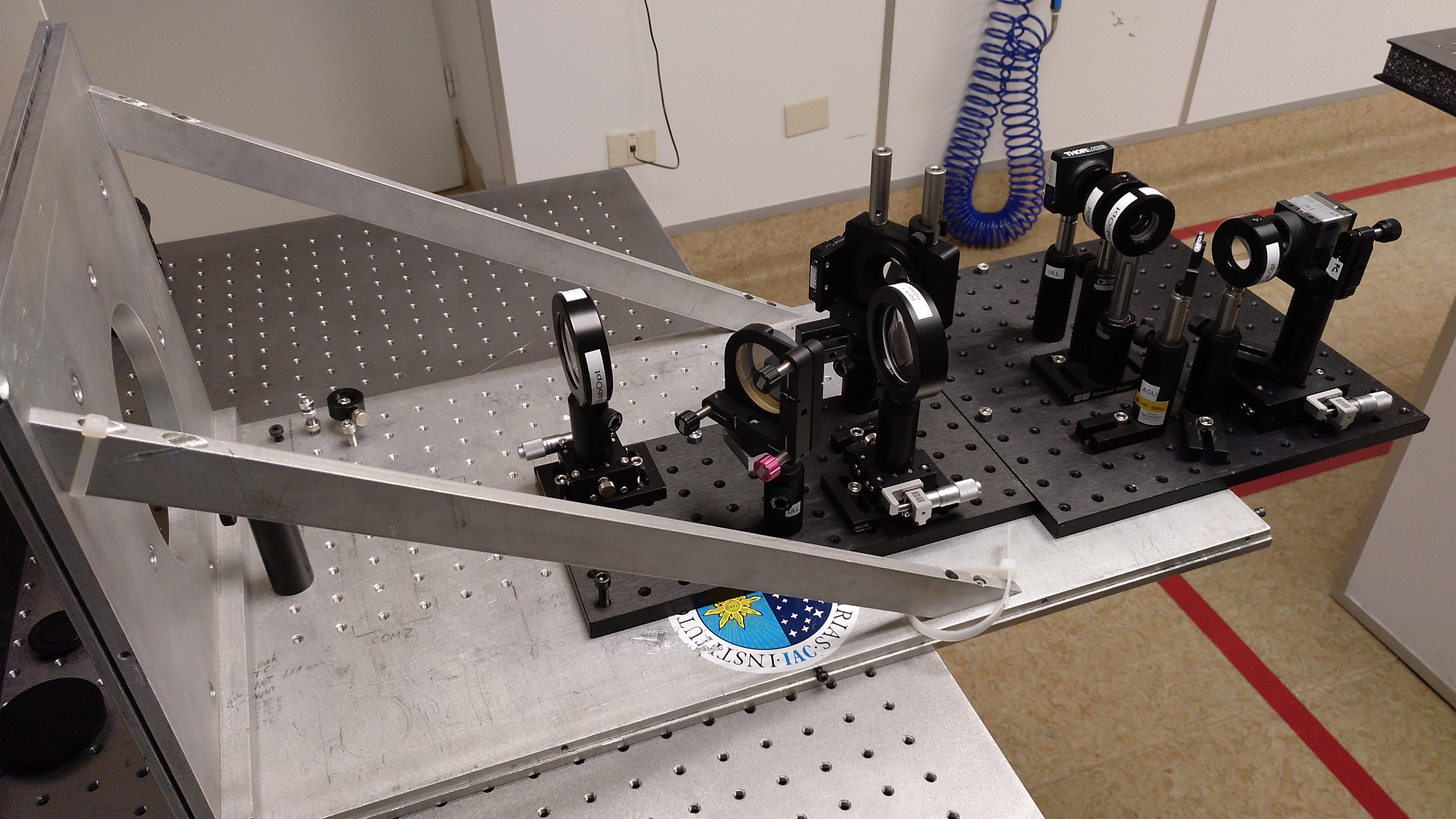}
\end{center}
\caption[foto] 
   { \label{fig:alioli_foto} 
Lateral view of the instrument mounted on the telescope interface for the TCS. Dummy cameras substitute the EMCCDs during the first development phase.}
\end{figure}

ALIOLI benefits from the new Two Pupil Plane Positions wavefront sensor (TP3-WFS) \cite{Colodro} and the Lucky Imaging techniques to achieve diffraction limited images through the Lucky Adaptive Optics. Thanks to the use of the TP3-WFS, the LI camera enhances the reachable resolution as it removes the highest scale turbulence maximizing the LI process. 

Currently on lab phase, we are exploring the performance of TP3-WFS vs traditional WFS such as Shack-Hartman and pyramid.

\begin{figure} [ht!]
\begin{center}
\includegraphics[height=6cm]{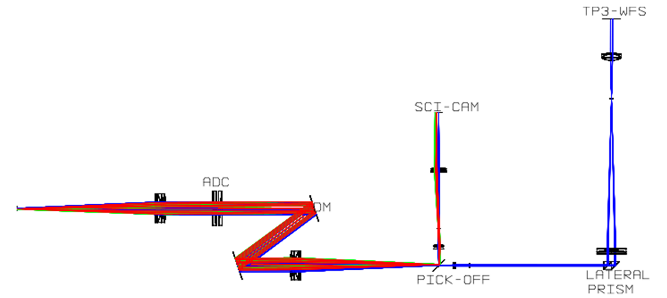}
\end{center}
\caption[layout] 
   { \label{fig:layout} 
ALIOLI optical layout with the TP3-WFS}
\end{figure} 

\section{THE KEY: a compact TP3-WFS}

\begin{figure} [ht!]
\begin{center}
\includegraphics[height=4cm]{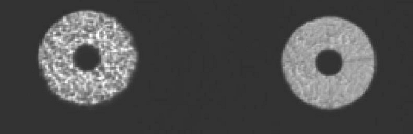}
\end{center}
\caption[pre_post] 
   { \label{fig:pre_post} 
The pre-pupil and post-pupil images of the TP3-WFS.}
\end{figure}

The clue to the ALIOLI instrument is the use of the TP3-WFS, it is a geometric WFS based on Van Dam\cite{Vandam} algorithms; it requires two defocused pupil planes, as seen on fig.\ref{fig:pre_post}. 

Unlike curvature wavefront sensors, the second derivative of the wavefront is not estimated. It assumes a purely geometrical propagation of the wavefront and calculates the first order derivatives directly. This assumption is only valid with short defocus distances, where diffraction has little effect on the images.

Supposed to be more sensitive than S-H sensors for low-order AO in the optical bands. AOLI is hence, the first instrument to succeed in closing the loop with stellar sources using the novel TP3-WFS. 

Working in visible band, represents an advance over classical wavefront sensors.

\section{LUcky AO}

The main idea is to perform AO driven frame selection. 

The Zernike modes needed to solve are only a few, for the WHT (4.2 m) only the first 14 are solved.

\begin{figure} [ht!]
	\begin{center}
		\begin{tabular}{c} 
			\includegraphics[height=8cm]{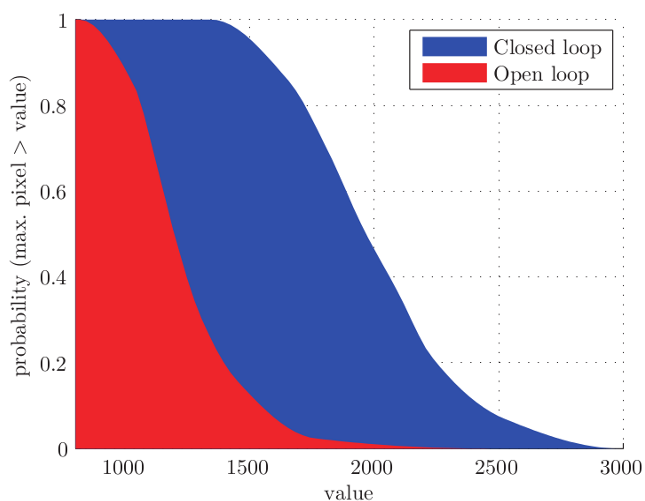}
			\end{tabular}
		\end{center}
		\caption[luckyao] 
		{ \label{fig:luckyao} 
		Probability of obtaining an image whose maximum
pixel value is higher than a given value, calculated from HIP10644.}
\end{figure}

Closing the AO loop gives a higher probability of getting an image of a given quality, meaning that more “lucky” frames are obtained, as seen in fig. \ref{fig:luckyao}

\bibliography{main} 
\bibliographystyle{spiebib} 

\end{document}